\documentclass[11pt]{article}\textwidth 6.5in\textheight 9in
\usepackage{amssymb}\usepackage[colorlinks]{hyperref}\usepackage{color}
	\usepackage{stmaryrd}\usepackage{mathrsfs}
	\usepackage{graphicx}
	\usepackage{amsmath}
	\usepackage{stmaryrd}
\usepackage{caption}

\begin{document}
\setlength{\captionmargin}{27pt}
\newcommand\hreff[1]{\href {http://#1} {\small http://#1}}
\newcommand\trm[1]{{\bf\em #1}} \newcommand\emm[1]{{\ensuremath{#1}}}
\newcommand\prf{\paragraph{Proof.}}\newcommand\qed{\hfill\emm\blacksquare}

\newtheorem{thr}{Theorem} 
\newtheorem{lmm}{Lemma}
\newtheorem{cor}{Corollary}
\newtheorem{con}{Conjecture} 
\newtheorem{prp}{Proposition}

\newtheorem{blk}{Block}
\newtheorem{dff}{Definition}
\newtheorem{asm}{Assumption}
\newtheorem{rmk}{Remark}
\newtheorem{clm}{Claim}
\newtheorem{example}{Example}

\newcommand{\ab}{a\!b}
\newcommand{\yx}{y\!x}
\newcommand{\yux}{y\!\underline{x}}

\newcommand\floor[1]{{\lfloor#1\rfloor}}\newcommand\ceil[1]{{\lceil#1\rceil}}

\newcommand{\lea}{<^+}
\newcommand{\gea}{>^+}
\newcommand{\eqa}{=^+}

\newcommand{\lel}{<^{\log}}
\newcommand{\gel}{>^{\log}}
\newcommand{\eql}{=^{\log}}

\newcommand{\llem}{\stackrel{\ast}{<}}
\newcommand{\gem}{\stackrel{\ast}{>}}
\newcommand{\eqm}{\stackrel{\ast}{=}}

\newcommand\edf{{\,\stackrel{\mbox{\tiny def}}=\,}}
\newcommand\edl{{\,\stackrel{\mbox{\tiny def}}\leq\,}}
\newcommand\then{\Rightarrow}

\newcommand\km{{\mathbf {km}}}\renewcommand\t{{\mathbf {t}}}
\newcommand\KM{{\mathbf {KM}}}\newcommand\m{{\mathbf {m}}}
\newcommand\md{{\mathbf {m}_{\mathbf{d}}}}\newcommand\mT{{\mathbf {m}_{\mathbf{T}}}}
\newcommand\K{{\mathbf K}} \newcommand\I{{\mathbf I}}
\newcommand\Kd{{\mathbf{Kd}}} \newcommand\KT{{\mathbf{KT}}} 
\renewcommand\d{{\mathbf d}} \newcommand\w{{\mathbf w}}
\newcommand\Ks{\Lambda} \newcommand\q{{\mathbf q}}
\newcommand\E{{\mathbf E}} \newcommand\St{{\mathbf S}}
\newcommand\M{{\mathbf M}}\newcommand\Q{{\mathbf Q}}
\newcommand\ch{{\mathcal H}} \renewcommand\l{\tau}
\newcommand\tb{{\mathbf t}} \renewcommand\L{{\mathbf L}}
\newcommand\bb{{\mathbf {bb}}}\newcommand\Km{{\mathbf {Km}}}
\renewcommand\q{{\mathbf q}}\newcommand\J{{\mathbf J}}
\newcommand\z{\mathbf{z}}

\newcommand\B{\mathbf{bb}}\newcommand\f{\mathbf{f}}
\newcommand\hd{\mathbf{0'}} \newcommand\T{{\mathbf T}}
\newcommand\R{\mathbb{R}}\renewcommand\Q{\mathbb{Q}}
\newcommand\N{\mathbb{N}}\newcommand\BT{\Sigma}
\newcommand\FS{\BT^*}\newcommand\IS{\BT^\infty}
\newcommand\FIS{\BT^{*\infty}}\newcommand\C{\mathcal{L}}
\renewcommand\S{\mathcal{C}}\newcommand\ST{\mathcal{S}}
\newcommand\UM{\nu_0}\newcommand\EN{\mathcal{W}}
\newcommand\W{\mathbb{W}}

\newcommand\lenum{\lbrack\!\lbrack}
\newcommand\renum{\rbrack\!\rbrack}

\renewcommand\qed{\hfill\emm\square}

\title{\vspace*{-3pc} The EL Theorem}

\author {Samuel Epstein\\
samepst@jptheorygroup.org}

\maketitle
\begin{abstract}
	The combined universal probability $\m(D)$ of strings $x$ in sets $D$ is close to max $\m(x)$ over $x$ in $D$: their logs differ by at most $D$'s information $\I(D:\ch)$ about the halting sequence $\ch$. 
\end{abstract}
\section{Introduction}
One common goal in computer science is to find the hidden part of the environment, this task has been called Inductive Inference, Extrapolation, Passive Learning, etc. The complete environment can represented as a huge string $x\in\FS$. The known observations restrict it to a set $D\subset \FS$. For example in thermodynamics, the environment $x$ can be seen as a record of every particle’s position and velocity in a closed box. An observation of some macro parameters, such as pressure and temperature, restricting the possible environments to a set $D$ of hypotheses consistent with the observation. \\

One method used to select a hypothesis (i.e. environment) is to leverage an \textit{ apriori} distribution over the environment space. This distribution $p$ encodes any knowledge about the environment known before the observation is made. Then selection of the hypothesis is 
$$\arg\max_{x \in D} p(x).$$

Note in AIT, for enumerable distributions  (i.e. generatable as outputs of randomized algorithms), there is a universal apriori distribution $\m(x)$. This is because $O(1)\m>p$, for all enumerable $p$. Furthermore, for all $x\in\FS$,  $\d(x|\m) = O(1)$, where $\d$ is deficiency of randomness; so there is no lower computable refutation to the statement: ``$x$ is generated from $\m$''. Thus when the universal prior is used, inductive  inference becomes an exercise of Occam’s razor:

$$\arg\min_{x \in D}\K(x).$$

However there exists a potential complication. It could be there is a collection $G\subset D$ of hypotheses representing a concept (such as a more detailed description of particles) where its combined apriori measure is greater than that of the simpliest element $x$, with $\m(x)\ll \m(G)$. Or, making the endeavor more murkier, it could be that $G$ is just the set of all complicated hypothesis and $G$ has greater combined apriori measure than the simpliest element. In this case, which explanation does one choose?\\

The EL Theorem shows that this dilemma is purely a mathematical construction. All the universal apriori measure of an observation $D$ is concentrated on its simpliest member. This is true for all non-exotic set $D$ with low mutual information with the halting sequence, $\I(D;\ch)$. There are no (randomized) algorithmic means of creating $D$ with arbitrarily high $\I(D;\ch)$.

\section{Related Work}
For information relating to the history of Algorithmic Information Theory and Kolmogorov complexity, we refer the readers to the textbooks~\cite{LiVi08} and \cite{DowneyHi10}. A survey about the shared information between strings and the halting sequence is in the work~\cite{VereshchaginVi04v2}. Work on the deficiency of randomness can be found in~\cite{Shen83,KolmogorovUs87,Vyugin87,Shen99}. Stochasticity of objects can be found in the works~\cite{Shen83,Shen99,Vyugin87,Vyugin99}. More information on stochasticity and algorithmic statistics are in the works \cite{GacsTrVi01,VereshchaginSh17,VereshchaginSh15}. The EL Theorem is joint work between the author and L. A. Levin who published this result in \cite{Levin16}.

\section{Conventions}
\label{sec:conv}
As noted in the introduction, $\K(x|y)$ is the conditional prefix free Kolmogorov complexity.  $\m(x)$ is the algorithmic probability. $\I(x;\ch)=\K(x)-\K(x|\ch)$ is the amount of information that the halting sequence $\ch\in\IS$ has about $x$. A probability is \textit{elementary}, if it has finite support and rational values. The deficiency of randomness of $x$ relative to a elementary probability measure $Q$ is $\d(x|Q)=-\log Q(x)-\K(x|Q)$. We recall for a set $D\subseteq\FS$, $\m(D)=\sum_{x\in D}\m(x)$. For the nonnegative real function $f$, we use $\lea f$, $\gea f$, and $\eqa f$ to denote $<f+O(1)$, $>f-O(1)$, and $=f\pm O(1)$. We also use $\lel f$ and $\gel f$ to denote $<f + O(\log (f+1))$ and $>f - O(\log (f+1))$, respectively. 

\section{The EL Theorem}
\label{sec:el}

\begin{dff}[Stochasticisty]
A string $x$ is $(\alpha,\beta)$-stochastic if there exists an elementary probability measure $Q$ such that
$$\K(Q)\leq \alpha \textrm{ and }\d(x|Q)\leq \beta.$$
\end{dff}
\begin{thr}[Epstein,Levin]
\label{thr:el}
Let $P$ be a lower-semicomputable semimeasure and $c$ be a large constant. Every $(\alpha,\beta)$-stochastic set $D$ with $s=\ceil{-\log P(D)}$ contains an element $x$ with
$$\K(x)<  s + \alpha + 2\log \beta + \K(s) +2\log \K(s)+c.$$
\end{thr}
The theorem is directly implied by the following lemma.
\begin{lmm}
\label{lmm:el}
Let $P$ be a lower-semicomputable semimeasure and $c$ be a large constant. If a set $D$ is $(\alpha,\beta)$-stochastic relative to an integer $s=\ceil{-\log P(D)}$, then $D$ contains an element $x$ with
$$
\K(x) < s + \alpha + \log \beta + \K(\log \beta) + \K(s)+c.
$$
Note that if $y$ is $(\alpha,\beta)$-stochastic relative to $s$, then it is $(\alpha,\beta+\K(s))$-stochastic. Hence the lemma implies the theorem.
\end{lmm}
\begin{lmm}
\label{lmm}
Let $P$ be a discrete mesure and $Q$ be a measure on sets. There exists a set $S$ of size $\ceil{\beta/\gamma}$ such that
$$
Q(\{ D : P(D)\geq\gamma \textrm{ and $D$ is disjoint from }S\})\leq \exp (-\beta).
$$
\end{lmm}
\begin{prf}
We use the probabilistic method, and show that if we draw $\ceil{\beta/\gamma}$ elements according to the distribution $P$, then the obtained set $S$ satisfies the inequality with positive probability. The probability that a fixed set $D$ with $P(D)\geq\gamma$ is disjoint from $S$ is
$$\leq (1-\gamma)^{\beta/\gamma}\leq \exp(-\beta).$$
Hence the expected $Q$-measure of such a $D$ is at most $\exp(-\beta)$ and the required set $S$ exists.\qed
\end{prf}$ $\\

\noindent\textbf{Proof of Lemma \ref{lmm:el} for computable $P$}. Let $Q$ be an elementary probability measure with $\K(Q)\leq \alpha$ and $\d(D|Q,s)\leq \beta$. Without loss of generality, we assume that $\beta$ is large positive power of 2. Fix a search procedure that on input $Q$, $\beta$, and $\gamma=2^{-s}$ finds a set satisfying the conditions of Lemma \ref{lmm}.\\

For large $\beta$, the set $D$ must intersect the obtained set $S$. Indeed, consider the $Q$-test $g(X|Q,s)$ that is equal to $\exp(\beta)$ if $X$ is disjoint from $S$, and is zero otherwise. This is indeed a test, because the above lemma implies that its expected value for $X\sim Q$ is bounded by 1. Since the test is also computable, it is a lower bound to the optimal test $\t(X|Q,s)$, up to a constant factor. By stochasticity of the set $D$, $g(D|Q,s)<O(1)\t(D|Q,s)<O(2^{\beta})$, because $2^{\d(X|Q,s)}$ is an optimal $Q$ test relative to $s$. Thus for large enough $\beta$, $D$ intersects $Q$.\\

It remains to construct a description of each element in $S$ of the size given in the proposition. We construct a special decompressor that assigns short description to each element in $S$. On input of a string, the decompressor interprets the string as a concatenation of 4 parts:
\begin{enumerate}
\item A prefix-free description of $Q$ of size at most $\alpha$.
\item A prefix-free description of $\log \beta$ of size $\K(\log \beta)$.
\item A prefix-free description of $s$ of size $\K(s)$.
\item An integer of bitsize $\log (\beta/\gamma) = s + \log \beta$. 
\end{enumerate}
It interprets the last integer as the index of an element in the set $S$ of size $\ceil{\beta/\gamma}$ that is computed by the search procedure on input $Q$, $\beta$, and $\gamma$. The element is the output of the decompressor. The proposition is proven for computable $P$.\qed

\begin{rmk}
If $P$ is computable, a set $S$ satisfying the conditions of the lemma can be easily searched. But if $P$ is not computable, then the collection of sets $D$ with $P(D) \geq\gamma$ grows over time. Thus after constructing a good S, it can happen that a large $Q$-measure of sets $D$ appears that does not contain an element from $S$, and that new elements to $S$ need to be added. This type of interactive construction leads to an equivalent characterization of the problem in terms of a game which is shown in \cite{Shen12}. Below, another proof is presented. 
\end{rmk}
\noindent\textbf{Proof of Lemma \ref{lmm:el} for lower-semicomputable $P$}. We still assume that $\beta$ is a large power of 2. Let $\gamma=2^{-s}/2$. We can rewrite $P=\frac{\gamma}{\beta}(P_1+\dots+P_f+P_*)$, with $f\leq \beta/\gamma$, such that $P_1,\dots P_f$ are probability measures with finite support obtained by a lower semi-computable approximation of $P$, and $P_*$ is a lower-semicomputable semimeasure.\\

\noindent\textit{Construction of a lower-semicomputable test $g$ over sets.} We first construct tests $g_1,\dots g_f$ together with a list of strings $z_1,\dots,z_f$. Let $g_0(X)=1$. Assume we already constructed $z_1,\dots,z_{i-1}$ and $g_{i-1}$ for some $i=1,\dots,f$. Choose $z_i$ such that the test
$$
g_i(X) = 
\left\{
\begin{array}{ll}
g_{i-1}(X)& \textrm{if }g_{i-1}(X)\geq \exp(\beta)\\
\exp(P_i(X))g_{i-1}(X)& \textrm{if }g_{i-1}(X)<\exp(\beta)\textrm{ and $X$ is disjoint from }\{z_1,\dots,z_i\}\\
0& \textrm{otherwise.}\\
\end{array}
\right.
$$
satisfies $\E g_i(X) \leq \E g_{i-1}(X)$ where the expectations are taken for $X\sim Q$. Let $g(X)$ be equal to $\exp(\beta)$ if there exists an $i$ such that $g_i(X)\geq\exp\beta$, otherwise let $g(X)=0$. \textit{End of construction}\\

We first show that each required string $z_i$ in the construction exists. Suppose $z_1,\dots,z_{i-1}$ and $g_{i-1}$ have already been constructed. We show the existence of $z_i$ using the probabilistic method. If we draw $z_i$ according to $P_i$, then for each set $X$ for which the second condition of $g_i$ is satisfied, we have
$$\E_{z_i\sim P_i} g_i(X)\leq (1-P_i(X))g_{i-1}(X)\exp P_i(X) \leq g_{i-1}(X),$$
because of the inequality $1+r\leq \exp(r)$ for all reals $r$. If $X$ satisfies the first or third condition, then $\E g_i(X)\leq \E g_{i-1}(X)$ is trivially true. So
\begin{align*}
\E_{X\sim Q}\E_{z_i\sim P_i}g_i(X)&\leq \E_{X\sim Q}g_{i-1}(X),\\
\E_{z_i\sim P_i}\E_{X\sim Q}g_i(X)&\leq \E_{X\sim Q}g_{i-1}(X),
\end{align*}
and the required $z_i$ exists.

We have $G(x)\leq O(\t(X|Q,(\gamma,\beta)))$, where $\t$ is the optimal test because the construction implies $\E g\leq 1$ and is effective, thus $g$ is lower semicomputable. Every set $X$ with $P(X)\geq 2^{-s}=2\gamma$ satisfies $P_1(X)+\dots+P_f(X)\geq\frac{\beta}{\gamma}P(D)-1\geq 2\beta -1 \geq \beta$ by choice of $P_i$. Any such $X$ that is disjoint from the set $\{z_1,\dots,z_f\}$ satisfies
$$
g_f(X) = \exp(P_1(X))\exp(P_2(X))\dots \exp(P_f(X))\geq \exp(\beta).
$$
This implies $\d(X|Q,s)>\beta$ for large $\beta$, because up to $O(1)$ constants, we have
$$
1.44\beta\leq \log g(X)\leq \d(X|Q,(\beta,\gamma))\leq \d(X|Q,s) + 2\log \beta.
$$
By the assumption on $(\alpha,\beta)$-stochasticity of $D$, we have $\d(D|Q,s)\leq \beta$ and hence $D$ must contain some $z_j$. The theorem follows by constructing a description for each string $z_i$ of bitsize $s+\alpha +\log\beta +\K(\log \beta) + \K(s)$ in a similar way as above.\qed

\subsection{Non-Stochastic Objects}
It is well known in the literature that non-stochastic objects have high mutual information with the halting sequence \cite{VereshchaginSh17}. In the following lemma, we reprove this fact, without using left-total machines, which was used in the original proof.\newpage
\begin{lmm}
	\label{lmm2}
	$\Ks(x)\lel \I(x;\ch)$.
\end{lmm}
\begin{prf}
	We dovetail all programs to the universal Turing machine $U$. For $p\in \mathrm{Domain}(U)$, $n(p)\in\N$ is the position in which the program $p\in\FS$ terminates. Let $\Omega^n = \sum_{p:n(p)< n}2^{-\|p\|}$ and $\Omega=\Omega^\infty$ be Chaitin's Omega. Let $\Omega^n_t$ be $\Omega^n$ restricted to the first $t$ digits. Let $x^*\in\BT^{\K(x)}$, with $U(x^*)=x$ with minimum $n(x^*)$. Let $k(p)=\max\{\ell:\Omega^{n(p)}_{\ell}=\Omega_{\ell}\}$ and $k=k(x^*)$. We define the elementary probability measure $Q(x) = \max\{2^{-\|p\|+k} : k(p)=k, U(p)=x\}$, $Q(\emptyset)=1-Q(\FS\setminus\{\emptyset\})$.
	\begin{align*}
	\d(x|Q) &= -\log Q(x) - \K(x|Q)\lea (\K(x) - k)- \K(x|\Omega_k)\\
	&\lea (\K(x|\Omega_k) + \K(\Omega_k) - k) - \K(x|\Omega_k)\lea (k + \K(k)) - k\\
	&\lea \K(k).\\ \\
	\K(x|\ch) &\lea\ \K(x|Q) + \K(Q|\ch)\lea  \K(x|Q) + \K(\Omega_k|\ch)\\
	&\lea -\log Q(x) + \K(k)\lea(\K(x) - k) + \K(k)\\
	k & \lel \K(x) - \K(x|\ch)\\ \\
	\Ks(x) &\lea \K(Q) + O(\log\max\{ \d(x|P),1\})\lea k + O(\K(k))\lel\I(x;\ch).
	\end{align*}\qed
\end{prf}
\begin{cor}[EL Theorem]$ $\\
	For finite $D\subset\FS$, $\min_{x\in D}\K(x) \lel -\log \m(D) + \I(D;\ch)$.
\end{cor}
\begin{prf}
This follows from Theorem \ref{thr:el} and Lemma \ref{lmm2}\qed
\end{prf}


\end{document}